\documentclass[10pt]{article}
\usepackage[english]{babel}
\usepackage{amsmath,amssymb,amsfonts}
\usepackage{subfigure}
\usepackage{multirow}
\usepackage{graphicx}
\usepackage[sectionbib]{natbib}
\usepackage{rotating}
\usepackage{setspace}
\usepackage{enumitem}
\usepackage{hyperref}

\doublespace

\usepackage[hmargin=3cm,vmargin=3.5cm]{geometry}

\setcitestyle{apa}

\title{Sequential sampling models in computational psychiatry:
Bayesian parameter estimation, model selection and classification}
\author{Thomas V. Wiecki}

\begin{document}

\maketitle

\begin{abstract}
Current psychiatric research is in crisis. In this review I will describe the causes of this crisis and highlight recent efforts to overcome current challenges. One particularly promising approach is the emerging field of computational psychiatry. By using methods and insights from computational cognitive neuroscience, computational psychiatry might enable us to move from a symptom-based description of mental illness to descriptors based on objective computational multidimensional functional variables. To exemplify this I will survey recent efforts towards this goal. I will then describe a set of methods that together form a toolbox of cognitive models to aid this research program. At the core of this toolbox are sequential sampling models which have been used to explain diverse cognitive neuroscience phenomena but have so far seen little adoption in psychiatric research. I will then describe how these models can be fitted to subject data and highlight how hierarchical Bayesian estimation provides a rich framework with many desirable properties and benefits compared to traditional optimization-based approaches. Finally, non-parametric Bayesian methods provide general solutions to the problem of classifying mental illness within this framework.
\end{abstract}

\newpage
\tableofcontents
\newpage

\section{Part I: Motivation}
Imagine going to a doctor because of chest-pain that has been bothering you for a couple of weeks. The doctor would sit down with you, listen carefully to your description of symptoms and prescribe medication to lower blood pressure in case you have a heart condition. After a couple of weeks your pain has not subsided. The doctor now prescribes medication against reflux which finally seems to help. In this scenario not a single medical analysis (e.g. EKG, blood work or a gastroscopy) was performed and medication with potentially severe side-effects prescribed on a trial-and-error basis. While highly unlikely to occur if you walked into a primary care unit with these symptoms today, this scenario is actually a realistic description if you had mental problems and had seen a psychiatrist instead.\\

There are several reasons for this discrepancy in sophistication between psychiatry and other fields of medicine. The main cause is that by definition, mental illness affects the brain -- the most complex biological system yet encountered. Compared to the level of scientific understanding achieved on other organs of the human body such as the heart, our understanding of the normally functioning brain is still, arguably, in its infancy.\\

Despite this complexity concerted efforts in the brain sciences have lead to an explosion of knowledge and understanding about the healthy and diseased brain in the last decades. The discovery of highly effective psychoactive drugs in the 50s and 60s raised expectations that psychiatry would progress in a similar fashion. Unfortunately, in retrospect it appears that these discoveries were serendipitous in nature as little progress has been made since then \citep[e.g][]{InselCuthbertGarveyEtAl10,Hyman12}. This lack of progress also caused many major pharmaceuticals companies like AstraZeneca and GlaxoSmithKline to withdraw from psychiatric drug development and close large research centers \citep{NuttGoodwin11,Cressey11}. In sum, psychiatry is a field in crisis \citep{PolandEckardtSpaulding94,InselCuthbertGarveyEtAl10,Hyman12,SahakianMallochKennard10}. As outlined in more detail below, the main reason for this crisis is a lack of measurable quantitative descriptors of mental illness. This lack results from an explanatory gap of how basic neurobiological aberrations result in complex disorders of the mind \citep{MontagueDolanFristonEtAl11,Hyman12}.\\

In part II I will review current challenges in psychiatry and recent efforts to overcome them. Several examples from the domain of decision making show the promise of moving away from symptom-based description of mental illness and instead formulate objective, quantifiable computational biomarkers as a basis for further psychiatric research. Part III introduces a computational cognitive toolbox that is suited to construct these computational biomarkers. Sequential sampling models serve as a case study for how computational models, when fit to behavior, have successfully been used to identify and quantify latent neurocognitive processes in healthy humans. Bayesian methods provide a resourceful framework to fit these models to behavior and establish individualized descriptors of neurocognitive function. After establishing the validity of these models to provide neurocognitive descriptors of individuals, I will review how clustering techniques can be used to construct a map of individual differences based on these neurocognitive descriptors. In sum, the objective of this review is to outline a research program to map the domain of neuropsychiatric disease.\\

In order to maintain a clear focus on this objective there are certain relevant issues not addressed here-within. While of critical importance to psychiatric patients, I will not discuss the clinical, pharmaceutical, environmental, social or developmental aspects of mental illness or rehabilitation programs. Moreover, I will treat mental illness as a disease of the brain with a focus on dysfunctional neurocircuitry \citep{Insel10}.



\subsection{Current challenges in psychiatry}
While the current crisis in psychiatry has complex causes that are deeply rooted in existing classification systems, one of the core problems is what has been identified by \citet{MontagueDolanFristonEtAl11} as the explanatory gap. This gap refers to our lack of understanding of the causal processes linking genes, molecular, cellular, neurocircuitry and cognition to psychiatric symptoms. This explanatory gap, coupled with the ``(almost) unreasonable effectiveness of psychotropic medication'' \citep{MontagueDolanFristonEtAl11} gave many a false premise to expect progress without understanding.

This explanatory gap suggests a new approach to research of mental disorders which aims to link cognitive and pure neuroscience to clinical symptoms without the restrictions of prior classification schemes \citep{PolandVon13,CuthbertInsel10,RobbinsGillanSmithEtAl12}. I will next specify problems current classification systems introduce followed by recent efforts to address them.

\subsubsection{Diagnostic and Statistical Manual of Mental Disorders}
For decades the Diagnostic and Statistical Manual of Mental Disorders (DSM) has been the basis of clinical diagnosis, treatment and research of mental illness. At its core, the DSM defines distinct disease categories like schizophrenia (SZ) and depression. These categories are mainly derived from translating subjective experience to objective symptomatology \citep{NordgaardSassParnas12} assuming unspecified biological, psychological, or behavioral dysfunctions \citep{PolandEckardtSpaulding94}.\\

While of certain value to clinicians the DSM is specifically designed to also serve as a classification system for scientific research with the goal of more easily translating results directly into clinical practice. While this translational research goal is commendable, decisions regarding systematic classification are more often based on perceptions of clinical utility rather than scientific merit \citep{PolandVon13}. As a consequence, DSM-based research programs failed to deliver consistent, replicable and specific results \citep{KendellJablensky03,Andreasen07,RegierNarrow09,KendlerKupferNarrowEtAl09,CuthbertInsel10,Hyman10}.\\

\paragraph{Heterogeneity and Comorbidity}
One major problem of contemporary psychiatry classification is the heterogeneus symptomatology of patients receiving identical diagnoses. One striking example of this is SZ where one must show at least 2 out of 5 symptoms to receive a diagnosis \citep{Heinrichs01}. It is thus possible to have patients with completely different symptomatology being diagnosed as schizophrenic.\\

Comorbidity is defined as the co-occurrence of multiple diseases in one individual. Importantly, we must differentiate between two relevant types of comorbidity: (i) True comorbidity is a result of independent disorders co-occurring; (ii) artificial comorbidity on the other hand is a result of separately classifying disorders that have a similar pathogenic cascade. It is now widely documented \citep{Markon10,KruegerMarkon06} that ``comorbidity between mental disorders is the rule rather than the exception, invading nearly all canonical diagnostic boundaries.'' \citep{BuckholtzMeyer-Lindenberg12}. The authors further note that ``It is important to understand that comorbidity in psychiatry does not imply the presence of multiple diseases or dysfunctions but rather reflects our current inability [to formulate] a single diagnosis to account for all symptoms.'' \footnote{See   \citet{BorsboomCramerSchmittmannEtAl11} for an alternative   explanation based on causal relationships between symptoms.} Together, these issues belie the assumption that DSM-disorders represent distinct and independent categories with a unique pathological cascade \citep{KruegerMarkon06,Hyman10,CuthbertInsel10,Andreasen07}.


\section{Part II: Potential Solutions}
As outlined above, the short-comings of the current DSM manual are well documented and a need for improvement has been recognized. In the following I will outline current efforts to address these challenges.

\subsection{Research Domain Criteria Project}
The Research Domain Criteria Project (RDoC) is an initiative by the National Institute for Mental Health (NIMH) \citep{InselCuthbertGarveyEtAl10}. RDoC improves on previous research efforts based on the DSM in the following ways. First, as the name implies it is conceptualized as a research framework only and is thus clearly separated from clinical practice. Second, RDoC is completely agnostic about DSM categories. Instead of a top-down approach which aims at identifying neural correlates of psychiatric disease, RDoC suggests a bottom-up approach that builds on the current understanding of neurobiological underpinnings of different cognitive processes and link those to clinical phenomena. Third, the RDoC research program integrates different levels of analysis like imaging, behavior and self-reports.\\

At its core, RDoC is structured into a matrix with columns representing different ``units of analysis'' and rows for research domains. The units of analysis include genes, molecules, cells, circuits, physiology, behavior, and self-reports. Research domains are clustered into negative and positive valence systems, cognitive systems, systems for social processes and arousal/regulatory systems. Each of these domains is further subdivided into distinct processes; for example, cognitive systems include attention, perception, working memory, declarative memory, language behavior and executive control.\\


Despite clear improvements over previous DSM-based research programs, the RDoC initiative currently lacks consideration of computational descriptors \citep{PolandVon13}. As I will outline below, computational methods show great promise to help link different levels of analysis, elucidate clinical symptoms and identify sub-groups of healthy and patient populations.

\subsection{Neurocognitive phenotyping}
In a recent review article, \citet{RobbinsGillanSmithEtAl12} suggest the use of neurocognitive endophenotypes to study psychiatric disease: ``Neurocognitive endophenotypes would furnish more quantitative measures of deficits by avoiding the exclusive use of clinical rating scales, and thereby provide more accurate descriptions of phenotypes for psychiatric genetics or for assessing the efficacy of novel treatments. The use of such measures would likely also facilitate and improve the use of informative animal models in psychiatry by focusing on cognitive and neural processes that can often be investigated in parallel across species. Defining such endophenotypes might cut across traditional psychiatric classification, and hence begin to explain the puzzle of apparent comorbidities.''\\

Of particular interest are three studies that use such neurocognitive endophenotypes by constructing dimensional functional profiles (MFPs) from summary statistics of a battery of various neuropsychological tasks to identify subtypes of ADHD \citep{DurstonFossellaMulderEtAl08,Sonuga-Barke05,FairBathulaNikolasEtAl12}.\\

\begin{figure}
\includegraphics{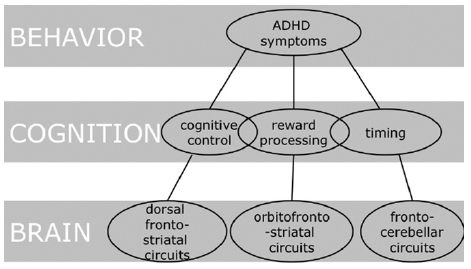}
\caption{Abnormalities in distinct brain areas (bottom level) can lead to different cognitive impairments (2nd level) and result in an ADHD diagnosis (top level). Figure reproduced from \citet{DurstonFossellaMulderEtAl08}.}
\label{durston}
\end{figure}

\citet{DurstonFossellaMulderEtAl08} argues that there are distinct pathogenic cascades along which abnormalities in at least three different brain circuits can lead to similar symptomatology. Specifically, abnormalities in dorsal frontostriatal, orbito-frontostriatal, or fronto-cerebellar circuits can lead to impairments of cognitive control, reward processing and timing, respectively. Core deficits in one or multiple of these brain networks can thus result in a clinical diagnosis of ADHD (see figure \ref{durston}) and provides a compelling explanation for the heterogeneity of the ADHD patient population. Preliminary evidence for this hypothesis is provided by \citet{Sonuga-Barke05} who used principal component analysis (PCA) on multi-dimensional functional profiles (based on a neuropsychological task battery) of ADHD patients and identified 3 distinct sub-types co-varying on timing, cognitive control, and reward.\\

\begin{figure}
\includegraphics[width=\columnwidth]{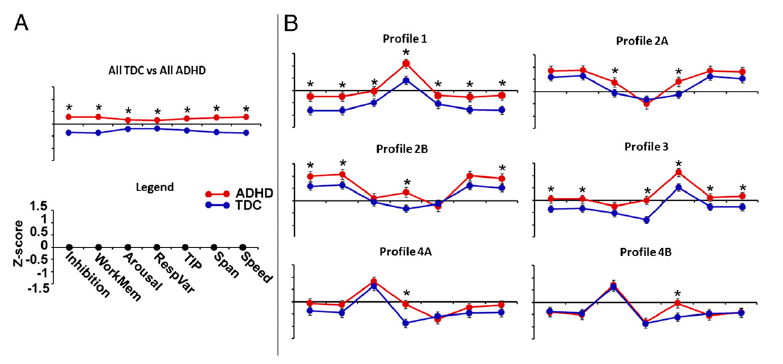}
\caption{Profile differences between ADHD and healthy controls (TDC) on based on functional descriptors. Lower left corner describes which each point along the x-axis represents. Upper left corner represents overall profile-differences. Right side shows differences between ADHD and HC functional profiles inside various clusters (i.e. profiles). As can be seen, group-differences vary over different profiles. Reproduced from \citet{FairBathulaNikolasEtAl12}.}
\label{Fair}
\end{figure}

A similar approach of identifying clusters in the ADHD population using MFPs was taken by \citet{FairBathulaNikolasEtAl12}. While similar in spirit to \citet{Sonuga-Barke05} and \citet{DurstonFossellaMulderEtAl08}, \citet{FairBathulaNikolasEtAl12} do not only look at differences in the patient population but in both, healthy controls (HC) and ADHD patients. The clustering is achieved by the application of graph theory. Interestingly, the authors find that HC and ADHD is not the predominant dimension along which clusters form. Instead, the authors uncover different functional profiles that apply to both, HC and ADHD patients. Critically, a classifier trained to predict HC and ADHD subjects inside of individual profiles achieved better performance than a classifier trained on the aggregated data. In addition, the abnormalities in cognitive functions of ADHD patients were different across different clusters (e.g. one cluster might show differences in response inhibition while another one shows differences in RT variability; see figure \ref{Fair}b). In other words, this implies that the overall population clusters into different cognitive profiles. ADHD affects individuals differently based on which cognitive profile they exhibit. Importantly, this study suggests that the source of heterogeneity may not only be distinct pathogenic cascades being labeled as the same disorder but may actually be a result of the inherent heterogeneity present in the overall population -- healthy {\it and} diseased. \\

The above mentioned studies all exemplify the danger of lumping subjects at the level of disease and treating them as one homogeneous category with a single, identifiable pathological cascade. Instead, these studies use MFPs to find an alternative characterization of subjects independent of their DSM classification that is (i) quantitatively measurable and (ii) a closer approximation to the underlying neurocircuitry \citep{RobbinsGillanSmithEtAl12}.

While a clear improvement on previous research efforts that use the DSM diagnosis as the sole descriptor this approach still has problems. First, although there is less reliance on DSM categories, these studies still use the diagnostic label for recruiting subjects. It could be imagined, for example, that patients with similar impulse control disorders like OCD or Tourette's have abnormalities in similar brain circuits; thus, if only OCD patients are recruited a critical part of the picture might be missed. Second, the cognitive task battery only covers certain aspects of cognitive function. Other tasks that for example measure working memory or reinforcement learning would be a useful addition. Finally, performance on each individual task is assessed by an aggregate performance score. Recent behavioral and neuropsychological findings, however, suggest that executive control in a single task may instead be more accurately characterized as a collection of related but separable abilities \citep{Baddeley96,ColletteVanLaureysEtAl05}, a pattern referred to as the “unity and diversity” of executive functions \citep{DuncanJohnsonFreer97,MiyakeFriedmanEmersonEtAl00}. Most cognitive tasks rely on a concerted and often intricate interaction of various neural networks and cognitive processes \citep[see e.g.][]{CollinsFrank12}. This task impurity problem \citep{Burgess97,Phillips97} complicates identification of separate brain circuits based solely on MFPs.\\

\begin{figure}
  \includegraphics{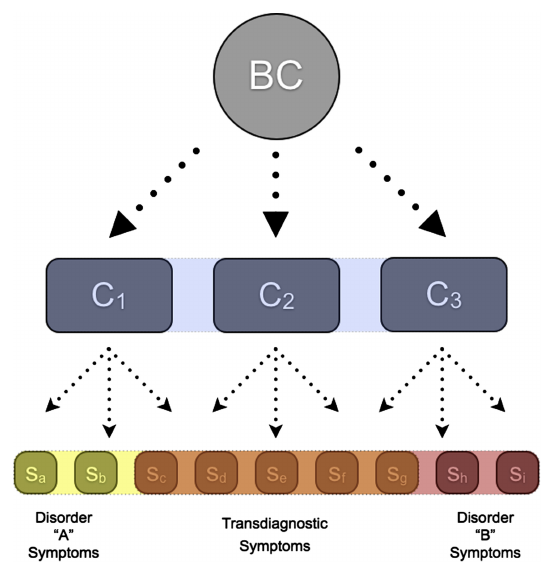}
\caption{Pathonegic cascade from Brain Circuit (BC) over multiple related cognitive processes (C1-C3) to symptoms (Sa-Si). ``Some of these symptoms will constitute diagnostic criteria for categorical disorder “A” but not disorder “B” (yellow shading), and some symptoms will be relatively selective for disorder “B” but not disorder “A” (red shading). However, the plurality of symptoms will overlap the two diagnostic categories (“transdiagnostic symptoms,” orange shading). This highlights the idea that connectivity circuits convey cognitive and symptom domain-specific, but disorder-general, genetic risk for mental illness.'' Reproduced from \citet{BuckholtzMeyer-Lindenberg12}.}
\label{Buckholtz}
\end{figure}

In sum, while cognitive phenotypes provide a useful framework for measuring brain function there is still ambiguity when using behavioral scores that present an aggregate measure of various brain networks. This issue is also discussed by \citet{BuckholtzMeyer-Lindenberg12} in relation to comorbidity: ``The fact that a brain circuit can be involved in multiple cognitive domains helps explain why diverse psychiatric disorders can exhibit common deficits and symptoms (comorbidity).'' This ``common symptom, common circuit'' model of psychopathology is illustrated by figure \ref{Buckholtz}. Disentangling these transdiagnostic patterns of psychiatric symptoms thus requires identification and measurement of underlying brain circuits. While the authors propose the use of functional imaging studies and genetic analysis I will discuss how computational modeling can contribute to disambiguate the multiple pathways leading to behavioral features.

\subsection{Computational psychiatry}
How have other brain sciences dealt with one-to-many mapping problem trying to dissociate behavior on a cognitive task with brain circuits? Computational models at different levels of abstraction have had tremendous impact on the field of cognitive neuroscience. The aim is to construct a model based on integrated evidence from neuroscience and psychology to explain neural activity as well as cognitive behavior. While more detailed biologically inspired models such as biophysical and neural network models are generally more constrained by neurobiology they often have many parameters which make it very difficult to fit them directly to human behavior. More abstract, process based models on the other hand often have fewer parameters that allow them to be fit directly to data at the cost of being less detailed about the neurobiology. Critically, all of these models allow for increased specificity in the identification of different neuronal and psychological processes that are often lumped together when analyzing task behavior based on summary statistics.\\

\begin{figure}
  \includegraphics[width=\columnwidth]{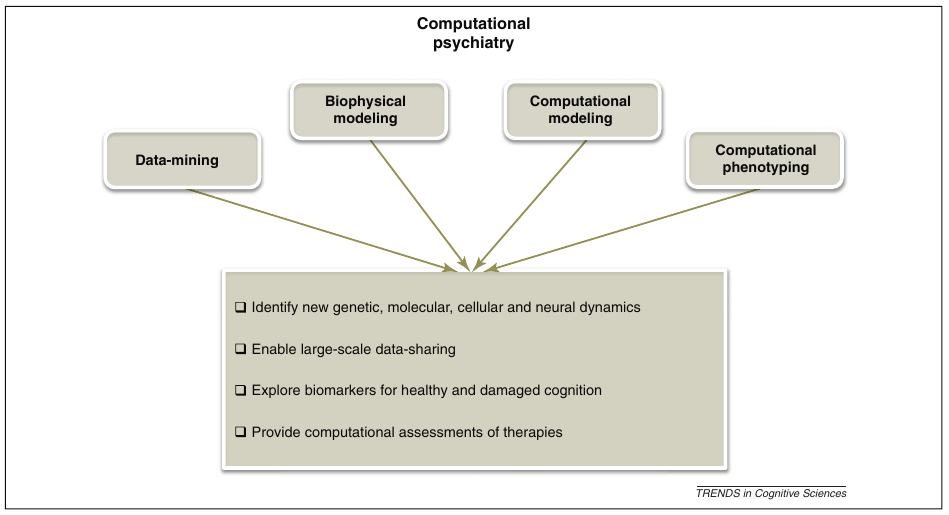}
  \caption{Overview of computational psychiatry. Different computational tools like computational modeling contribute to insights in mental health research. Reproduced from \citet{MontagueDolanFristonEtAl11}.}
  \label{comp_psyc_montague}
\end{figure}

Using computational models to infer dysfunctional latent processes in the brain is the premise of the newly emerging field of computational psychiatry. In their groundbreaking review, \citet{MontagueDolanFristonEtAl11} define the goal for computational psychiatry of ``extracting computational principles around which human cognition and its supporting biological apparatus is organized. Achieving this goal will require new types of phenotyping approaches, in which computational parameters are estimated (neurally and behaviorally) from human subjects and used to inform the models. This type of large-scale computational phenotyping
of human behavior does not yet exist.'' Also, see figure \ref{comp_psyc_montague} for a visual depiction.\\

\begin{figure}
\includegraphics{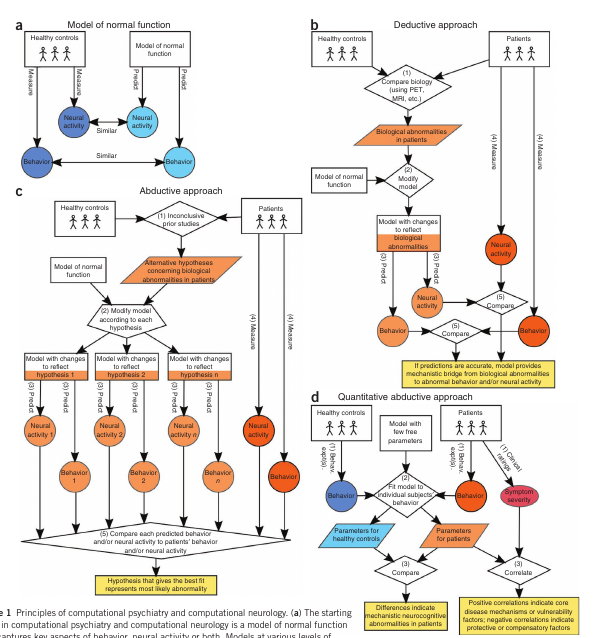}
\caption{Different approaches computational models can inform mental health research. Given a computational model of normal function (a), research can provide a mechanistic bridge from neural abnormalities to explain or compare behavioral differences in a deductive approach (b). Contrary, in an abductive approach (c) behavioral differences are used to infer underlying neuronal abnormalities. Similarly, computational models can be quantitatively fit to behavior to infer underlying causes (i.e. quantitative abductive approach; d). See text for more information. Reproduced from \citet{MaiaFrank11}.}
\label{comp_psyc_maia}
\end{figure}

Based on this premise, \citet{MaiaFrank11} identify computational models as a ``valuable tool in taming [the complex pathological cascades of mental illness] as they foster a mechanistic understanding that can span multiple levels of analysis and can explain how changes to one component of the system (for example, increases in striatal D2 receptor density) can produce systems-level changes that translate to changes in behavior''. Moreover, the authors define three concrete strategies for how computational models can be used to study brain dysfunction (see also figure \ref{comp_psyc_maia}) given a model of normal function:
\begin{itemize}
\item {\it Deductive approach}: Established neuronal models can be tested for how pathophysiologically plausible alterations in connectivity or neurotransmitter levels (e.g. dopamine is known to be reduced in Parkinson's disease) affect system level activations and behavior. This is essentially a bottom-up approach as it involves the study of how known or hypothesized neuronal changes affect higher-level functioning.
\item {\it Abductive approach}: Computational models can be used to infer neurobiological causes from known behavioral differences. In essence, this is a top-down approach which tries to link behavioral consequences back to underlying latent causes.
\item {\it Quantitative abductive approach}: Parameters of a computational model are fit to a subjects' behavior on a suitable task or task battery. Different parameter values point to differences in underlying neurocircuitry of the associated subject or subject group. These parameters can either be used comparatively to study group differences (e.g. healthy and diseased) or as a regressor with e.g. symptom severity. This approach is more common with abstract models than with neural network models as the former have often fewer parameters and thus can be more easily fit to data.
\end{itemize}

\subsubsection{Case studies in the domain of decision making}
One key area in which computational models have had tremendous success in elucidating how the different cognitive and neurobiological gears work together is the domain of decision making. In addition, many mental illnesses can be characterized by aberrant decision making of one sort or another \citep{MaiaFrank11,WieckiFrank10,MontagueDolanFristonEtAl11}. In the following I will review recent cases where computational models of
decision making have been used to better understand brain disorders.\\

\paragraph{Computational models of reinforcement learning}
\subparagraph{Parkinson's Disease}
Our first case study concerns Parkinson's disease (PD). Its most visible symptoms affect the motor system as manifest as hypokinesia, bradykinesia, akinesia, rigidity, tremor and progressive motor degeneration. However, recently, cognitive symptoms have received increased attention \citep[e.g.,][]{Cools06,Frank05,MoustafaShermanFrank08,DaCunhaWietzikoskiDombrowskiEtAl09}. PD is an intriguing neuropsychiatric disorder because its pathogenic cascade is well identified to be the cell death of midbrain dopaminergic neurons in the substantia nigra pars compacta (SNc), part of the basal ganglia (BG) \citep{KishShannakHornykiewicz88}. Neural network models of the BG \citep{Frank05,Frank06} interpret this brain network as an adaptive action selection device that conditionally gates internal or external actions based on their previous reward history. DA is critically involved in learning from rewards and punishments which actions to facilitate and which actions to suppress in the future \citep{LjungbergApicellaSchultz92,MontagueDayanSejnowski96,Schultz98,WaeltiDickinsonSchultz01,PanSchmidtWickensEtAl05,BayerLauGlimcher07,RoeschCaluSchoenbaum07,SuttonBarto90,Barto95,MontagueDayanSejnowski96,SchultzDayanMontague97}. Behavioral reinforcement learning tasks show that the chronic low levels of DA in PD patients result in a bias towards learning from punishment at the cost of learning from rewards \citep{FrankSeebergerOReilly04,CohenFrank09}. In extension, we have argued that PD is not a motor disorder per se but rather an action selection disorder in which the progressive decline of motor and cognitive function can be interpreted in terms of aberrant learning \textit{not} to select actions
\citep{WieckiFrank10,WieckiRiedingerAmeln-MayerhoferEtAl09}.\\

In this case study, an existing biological model of normative brain
function was paired with a known and well localized neuronal
dysfunction to extend our understanding of the symptomatology of a
brain disorder. Note, however, that the model was not fit to data
quantitatively. In the terminology established by \citet{MaiaFrank11},
this is an example of the {\it deductive} approach in which the model
provides a mechanistic bridge that explains how abnormal behavior can result from neurocircuit dysfunctions.

\begin{figure}
\includegraphics[width=\columnwidth]{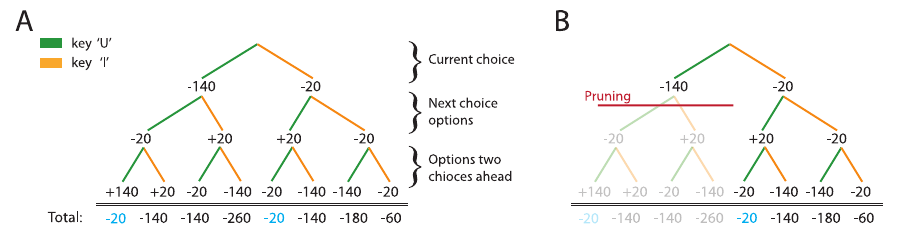}
\caption{Depiction of decision trees with 2 possible actions (pressing of 'U' key (green) or 'I' key (yellow). Each action (starting at the top node) leads to winning or losing points and subsequent action choice. After 3 subsequent actions points are aggregated. Critically, subjects experience rewards and have to learn the reward structure (which remains stable across a number of trials) to maximize rewards. Complete enumeration of a decision tree (left) has exponential complexity and is thus computationally infeasible. Thus, the decision tree must be pruned at a certain point. One possible strategy is to prune once a large negative reward is encountered (as depicted). Reproduced from \citet{HuysEshelONionsEtAl12}.}
\label{pruning}
\end{figure}

\subparagraph{Depression}
Our second case study involves how depression may be understood as an action planning disorder. Planning a series of future actions is an exponentially complex problem as each individual action can have different outcomes which themselves enable new actions essentially forming a decision tree. One way to deal with this complexity is to prune the decision tree and not consider certain actions (see figure \ref{pruning}). Recently, \citet{HuysEshelONionsEtAl12} proposed a model of how humans perform this approximation. Briefly, the authors suggest that actions that would lead to comparatively bad outcomes will not be further considered and pruned from the decision tree (note that this is provably not optimal as early bad outcomes could be more than accounted for later on). To test this theory the authors test healthy subjects on a novel behavioral task that requires execution of a sequence of actions each associated with winning or losing a certain number of points. Critically, in one task condition, subjects had two action paths available to them: (i) one path in which a very large loss occurred early that, however, was more than accounted for later on; and (ii) another less favorable one action path without an early loss but lower cumulative rewards overall. Participants overwhelmingly chose the non-optimal action sequence that did not result in an early large loss suggesting that humans prune the decision tree once actions are encountered leading to unfavorable outcomes, even in scenarios where this strategy does not result in optimal performance.\\

The authors formulate, fit and compare various computational algorithmic models with different pruning strategies. The model that provided the best fit (measured with an integrated BIC score that penalizes model complexity; see below) had three parameters: specific pruning, general pruning and reward sensitivity. Specific pruning represents the probability of a participant to stop evaluating sub-trees that are associated with large losses while general pruning represents the probability to stop considering sub-trees irrespective of rewards.  Reward sensitivity captured the tendency to evaluate a loss of e.g.  -140 to be of larger absolute magnitude than a win of 140 points although they cancel each other (i.e. loss aversion).\\

\begin{figure}
\center
\includegraphics{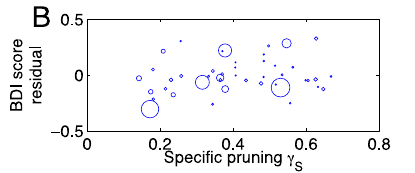}
\caption{Correlation between specific pruning parameter $\lambda_S$ and residual depression rating (BDI). Each circle represents one subject. Diameter of the circle corresponds to estimation uncertainty. See text for further model details. Reproduced from \citet{HuysEshelONionsEtAl12}.}
\label{bdi}
\end{figure}

Participants were also given questionaires to assess their
(sub-clinical) levels of depression (BDI). Intriguingly, this
depression rating correlated with the specific pruning parameter of
the model (see figure \ref{bdi}). This effect was specific to this
parameter as there was no such correlation with the general pruning
parameter. In other words, subjects with higher depressive ratings
were quicker to discard plans that lead to bad outcomes early on --
sometimes missing large rewards available to them later on. Finally,
the authors speculate that the specific pruning depends on the
serotonergic transmitter system.\\

In sum, \citet{HuysEshelONionsEtAl12} proposed an algorithmic model of
normative cognitive computation. By quantitatively fitting the model
to behavioral data on a novel task and regressing an independent
clinical variable (rating on the depression scale) with the fitted
parameter values the authors are able pinpoint the cognitive
computation underlying a clinically significant symptom. In terms of
the \citet{MaiaFrank11} terminology, \citet{HuysEshelONionsEtAl12}
used a {\it quantitative abductive} approach.

\subparagraph{Schizophrenia} Despite SZ being the focus of intense research over the last decades, no single theory of its underlying neural causes has been able to explain the diverse set of symptoms that can lead to a SZ diagnosis.  The symptomatology is structured in terms of positive symptoms like psychosis, negative symptoms like anhedonia which refers to the inability to experience pleasure from activities usually found enjoyable such as social interaction. \\

Recent progress has been made by the application of RL models to understand individual symptoms or a single symptom category (e.g. negative symptoms) rather than SZ as a whole \citep{WaltzFrankWieckiEtAl11,GoldWaltzPrenticeEtAl08,GoldWaltzMatveevaEtAl12,StraussFrankWaltzEtAl11}.

In a recent behavioral study using a RL task, \citet{WaltzFrankRobinsonEtAl07} found that SZ patients show reduced performance in selecting previously rewarded stimuli compared with HCs. Moreover, this performance deficit is most pronounced in patients with severe negative symptoms. Notably, SZ and HC did not differ in their ability to avoid actions leading to negative outcomes. However, this behavioral analysis did not allow to differentiate whether SZ patients were impaired at learning from positive outcomes or from a failure in representation of the prospective reward values during
decision making.

This dichotomy in learning vs representation is also present in two types of RL models -- actor-critic and Q-learning models \citep{SuttonBarto98}. An actor-critic model consists of two modules: an actor and a critic. The critic learns the expected rewards of states and trains the actor to perform actions that lead to better-than expected outcomes. Q-learning models on the other hand have an explicit representation of the outcomes that are associated with each action. Thus, while a Q-learning model chooses actions based on their absolute reward values, an actor-critic chooses actions based on whether they lead to better-than-expected outcomes.

In a follow-up study, \citet{GoldWaltzMatveevaEtAl12} administered a new task that paired a neutral stimulus in one context with a positive and in another context with a negative stimulus. While the neutral stimulus has the same value of zero in both contexts, it is known that DA signals reward prediction errors (RPE) that drive learning in the BG are coding outcomes {\it relative} to the expected reward \citep{SuttonBarto90,Barto95,MontagueDayanSejnowski96,SchultzDayanMontague97}. Thus, in the negative context, receiving nothing is better than expected and will result in a positive RPE, driving learning in the BG to select this action in the future \citep{Frank05}. In a test period in which no rewards were presented, participants had to choose between an action that had been rewarding and one that had simply avoided a loss. Both actions should have been associated with better-than-expected outcomes. An actor-critic model should thus show a tendency to select the neutral stimulus while a Q-learning model with explicit representation of the reward contingencies should mainly select the one with a higher reward. Intriguingly, when both of these models were fit to participant data, the actor-critic model produced a better fit for SZ patients with high degree of negative symptoms while HC and SZ with low negative symptoms were better fit by a Q-learning model. In other words, the high negative symptoms group largely based decisions on learned stimulus-response associations instead of expected reward values. Notably, HC and the low negative symptom group did not differ significantly in their RL behavior. Moreover, this also rules out possibly confounding effects of antipsychotic drugs, as both patient groups were similarly medicated.\\

In a related line of work, \citet{StraussFrankWaltzEtAl11} tested HC and SZ patients on a reinforcement learning task that allowed subjects to either adopt a safe strategy and {\it exploit} the rewards of actions with previously experienced rewards, or, to {\it explore} new actions with perhaps even higher payoffs. \citet{FrankDollOas-TerpstraEtAl09} develop a computational model that can be fit to subjects' behavior and recover how individual subjects balance this exploration-exploitation trade-off. Intriguingly, applying this model to SZ patients, \citet{StraussFrankWaltzEtAl11} found that patients with high anhedonia ratings where less willing to explore their environment and uncover potentially better actions. This result suggests a reinterpretation of the computational cognitive process underlying anhedonia. For example, one might assume that the lack of engagement of social activities of anhedonistic patients results from an inability to experience pleasure and as a consequence, a failure to learn the positive value of social interaction. Instead, this study suggests that anhedonia is a result of an inability to consider the prospective benefit of doing something that might lead to better outcomes.\\

In sum, \citet{GoldWaltzMatveevaEtAl12} and \citet{StraussFrankWaltzEtAl11} used a {\it quantitative abductive} approach to infer aberrant computational cognitive processes in RL in a subgroup of SZ patients. By grouping subjects according to symptom severity instead of diagnosis the authors addressed the problem of heterogeneity. Moreover, linking the results to prior neural network modeling efforts \citep{FrankClaus06}, as described above, also points to the OFC as a promising target for further investigation as a neural source of negative symptoms in SZ patients.

\paragraph{Computational models of response inhibition}
Besides RL, response inhibition is another widely studied phenomenon
in cognitive neuroscience of relevance to mental illness. Response
inhibition is required when actions in the planning or execution stage
are no longer appropriate and must be suppressed. The antisaccade task
is one such task that is often used in a psychiatric setting \citep[e.g][]{AichertWostmannCostaEtAl12,Fukumoto-MotoshitaMatsuuraOhkuboEtAl09}. It requires subjects to inhibit a prepotent response to a salient
stimulus and instead saccade to the opposite side \citep{Hallett78}. A wealth of literature has demonstrated reduced performance of psychiatric patients with disorders including
attention\-deficit/hyper\-activity disorder (ADHD) \citep{Nigg01,OosterlaanLoganSergeant98,SchacharLogan90}, obsessive
compulsive disorder (OCD)
\citep{ChamberlainFinebergBlackwellEtAl06,MenziesAchardChamberlainEtAl07,PenadesCatalanRubiaEtAl07,Morein-ZamirFinebergRobbinsEtAl09},
schizophrenia (SZ)
\citep{HuddyAronHarrisonEtAl09,BellgroveChambersVanceEtAl06,BadcockMichieJohnsonEtAl02},
Parkinson's disease (PD) \citep{van_KoningsbruggenPenderMachadoEtAl09}
and substance abuse disorders
\citep{MonterossoAronCordovaEtAl05,NiggWongMartelEtAl06}. However, as
demonstrated by \citet{WieckiFrank13}, even a supposedly simple
behavioral task such as the antisaccade task requires a finely
orchestrated interplay between various brain regions including frontal
cortex and basal ganglia. It thus can not be said that
decreased accuracy in this task is evidence of response inhibition
deficits per se as the source of this performance impairment can be
manifold.\\

In follow-up work, we have formulated a psychological process model which summarizes the higher-level computations of the neural network and has fewer parameters. In an attempt to bridge these two levels of abstraction, we fit this process model to the outputs of this neural network model for which the biological modulations can be tightly controlled \citep{WieckiFrank10}. Interestingly, by modulating different biological parameters in the neural network model and recovering which parameter of the process model was affected by this modulation, we were able to associate high-level computational processes with their neural correlates. The hope is that these associations, once validated, allow us to infer specific neural aberrations from behavioral performance.\\

In sum, computational models like the DDM allow mapping of behavior to
psychological processes and could thus be categorized as the {\it computational abductive} approach. However, ambiguity of how
psychological processes relate to the underlying neurocircuitry still
have to be disambiguated. By combining different levels of modeling
these ambiguities can be better identified and studied. Ultimately,
this might allow development of tasks that use specific conditions
(e.g. speed-accuracy trade-off, reward modulations and conflict) to
disambiguate the mapping of psychological processes to their
neurocircuitry. Using biological process models to test different
hypotheses about the behavioral and cognitive effects of neurocircuit
modulations would correspond to the {\it deductive} approach. In other
words, by combining the research approaches outlined by
\citet{MaiaFrank11} we can use our understanding of the different
levels of processing to inform and validate how these levels interact
in the healthy and dysfunctional brain.\\

In sum, there are a few example studies which applied established computational models to identify model parameters (which aim to describe a single cognitive function) and relate it to the severity of a specific clinical symptom. In the following, I will review sequential sampling models and Bayesian methods and show how they are well suited to solve many of the issues encountered in computational psychiatry.

\section{Part III: Quantitative Methods}
Part I described the current issues in psychiatry. Several examples presented in part II highlighted computational psychiatry as an area with a lot of potential to solve these problems. In part III I will review several quantitative methods that can solve the problems associated with quantifying cognitive function. Specifically, sequential sampling models present a versatile tool to model cognitive function. Fitting these models to data -- especially with the small number of trials often found in clinical experiments -- is a challenge well addressed by hierarchical Bayesian models that share statistical power by assuming similarity between subjects. However, there are two limitations with this basic approach. (i) Traditional inference methods require a likelihood function which is often intractable for more nuanced formulations of sequential sampling models. Likelihood free methods solve this problem as they only require a generative process from which the likelihood is estimated. (ii) While the similarity assumption made by hierarchical Bayesian modeling is reasonable we can not know the exact form of this similarity ahead of time. Bayesian mixture models address this problem by inferring clusters from the data. While traditional methods like Gaussian Mixture Models require specification of the number of clusters to be found in the data ahead of time, Bayesian non-parametrics relax this restriction and infer the number of clusters from the data. In the following I will focus on how each quantitative method helps to solve the above mentioned issues. Mathematical details can be found in the appendix. Finally, while these methods are described with the motivation of estimating sequential sampling models, the Bayesian methods are applicable more broadly (e.g. to the estimation of RL models).

\subsection{Sequential Sampling models}
Cognition spans many mental processes that include attention, social cognition, memory, emotion, and reasoning, to name a few. As outlined above, RL models have already proven to be a valuable tool in explaining neuropsychological disorders and their symptoms. A computational psychiatric framework that aims to explain the multi-faceted domain of mental illness must thus include computational cognitive neuroscience models that cover a broad range of cognitive processes. As such a review is outside of our scope (but see \citet[e.g.][]{OReillyMunakata00}) I will focus on sequential sampling models as an illustrative example for how these models work, how they have been applied to study normal and aberrant neurocognitive phenomena, how they can be fit to data using Bayesian estimation and how subgroups of subjects can be inferred using mixture models.\\

Sequential sampling models (SSMs) \citep[e.g.][]{TownsendAshby83} have established themselves as the de-facto standard for modeling data from simple decision making tasks \citep[e.g.][]{SmithRatcliff04}. Each decision is modeled as a sequential extraction and accumulation of information from the environment and/or internal representations. Once the accumulated evidence crosses a threshold, a corresponding response is executed. This simple assumption about the underlying psychological process has the intriguing property of reproducing reaction time distributions and choice probability in simple two-choice decision making tasks.\\

SSMs generally fall into one of two classes: (i) diffusion models which assume that {\it relative} evidence is accumulated over time and (ii) race models which assume independent evidence accumulation and response commitment once the first accumulator crossed a boundary \citep[e.g.][]{LaBerge62,Vickers70}. While there are many variants of these models they are often closely related on a computational level and sometimes mathematically equivalent under certain assumptions \citep{BogaczBrownMoehlisEtAl06}. As such, I will restrict discussion to two exemplar models from each class widely used in the literature: the drift diffusion model (DDM) \citep{RatcliffRouder98,RatcliffMcKoon08} belonging to the class of diffusion models and the linear ballistic accumulator (LBA) \citep{BrownHeathcote08} belonging to the class of race models.

\subsubsection{Drift Diffusion Model}
The DDM models decision making in two-choice tasks. Each choice is represented as an upper and lower boundary. A drift-process accumulates evidence over time until it crosses one of the two boundaries and initiates the corresponding response \citep{RatcliffRouder98,SmithRatcliff04}. The speed with which the accumulation process approaches one of the two boundaries is called the drift rate and represents the relative evidence for or against a particular response. Because there is noise in the drift process, the time of the boundary crossing and the selected response will vary between trials. The distance between the two boundaries (i.e. threshold) influences how much evidence must be accumulated until a response is executed. A lower threshold makes responding faster in general but increases the influence of noise on decision making while a higher threshold leads to more cautious responding. Reaction time, however, is not solely comprised of the decision making process -- perception, movement initiation and execution all take time and are summarized into one variable called non-decision time. The starting point of the drift process relative to the two boundaries can influence if one response has a prepotent bias. This pattern gives rise to the reaction time distributions of both choices (see figure \ref{DDM}; mathematical details can be found in the appendix).\\

\begin{figure}
\includegraphics[width=\columnwidth]{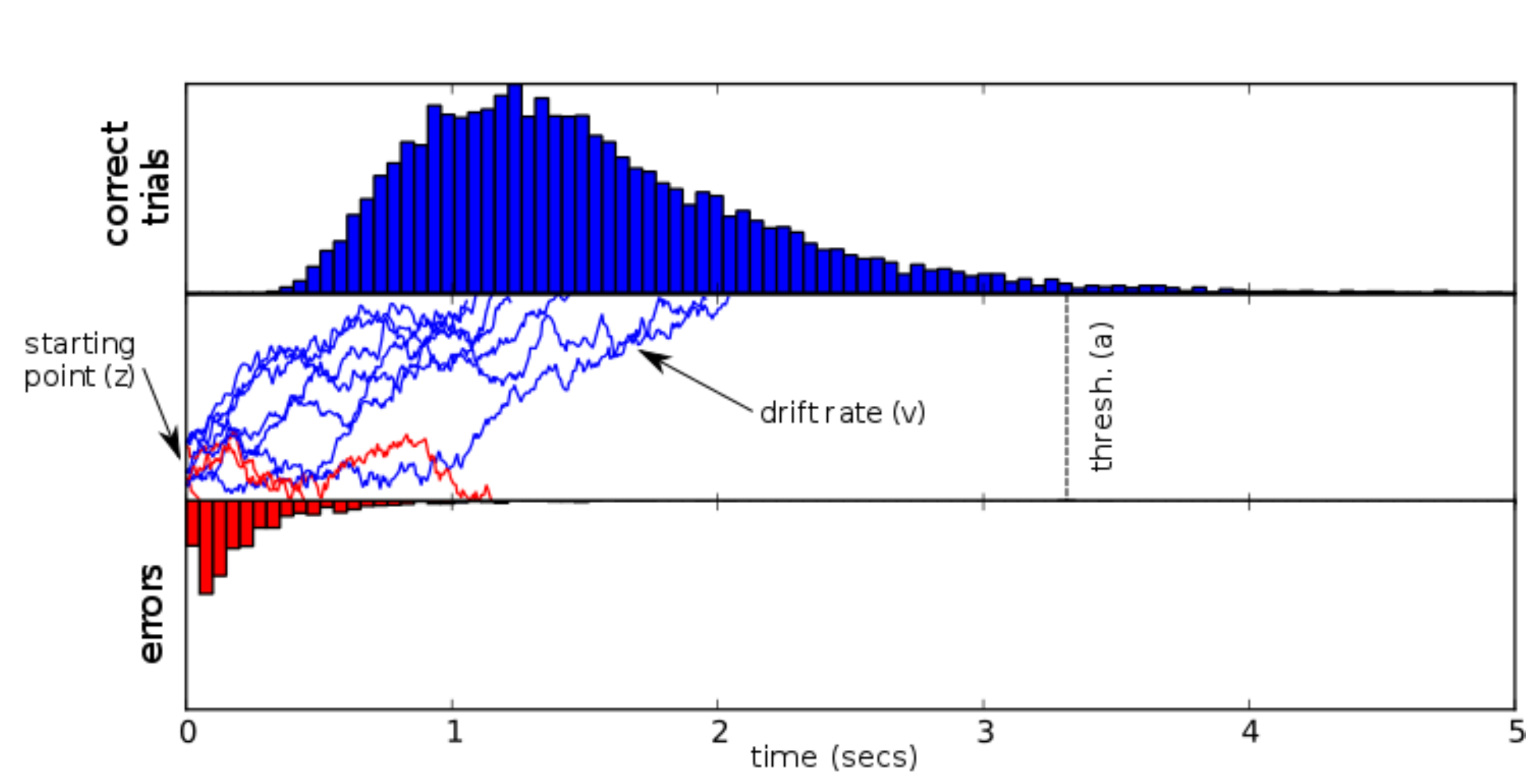}
\caption{Trajectories of multiple drift-processs (blue and red lines, middle panel). Evidence is accumulated over time (x-axis) with drift-rate v until one of two boundaries (separated by threshold a) is crossed and a response is initiated. Upper (blue) and lower (red) panels contain histograms over boundary-crossing-times for two possible responses. The histogram shapes match closely to that observed in reaction time measurements of research participants.}
\label{DDM}
\end{figure}

Later on, the DDM was extended to include inter-trial variability in the drift-rate, the non-decision time and the starting point in order to account for two phenomena observed in decision making tasks -- early and late errors. Models that take this into account are referred to as the full DDM \citep{RatcliffRouder98}.

\subsubsection{Linear Ballistic Accumulator}
\begin{figure}
\includegraphics[width=\columnwidth]{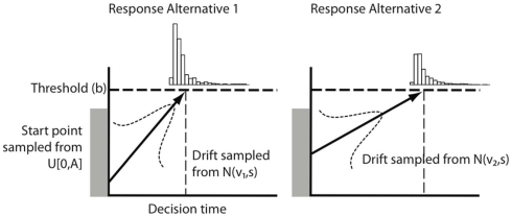}
\caption{Two linear ballistic accumulators (left and right) with different noiseless drifts (arrows) sampled from a normal distribution initiated at different starting points sampled from uniform distribution. In this case, accumulator for response alternative 1 reaches criterion first and gets executed. Because of this race between two accumulators towards a common threshold these model are called race-models. Reproduced from \citet{DonkinBrownHeathcoteEtAl11}.}
\label{LBA}
\end{figure}

The Linear Ballistic Accumulator (LBA) model belongs to the class of race models \citep{BrownHeathcote08}. Instead of one drift process and two boundaries, the LBA contains one drift process for each possible response with a single boundary each. Thus, the LBA can model decision making when more than two responses are possible. Moreover, unlike the DDM, the LBA drift process has no intra-trial variance. RT variability is obtained by including inter-trial variability in the drift-rate and the starting point distribution (see figure \ref{LBA}). Note that the simplifying assumption of a noiseless drift-process simplifies the math significantly leading to a computationally faster likelihood function for this model.

In a simulation study it was shown that the LBA and DDM lead to similar results as to which parameters are affected by certain manipulations \citep{DonkinBrownHeathcoteEtAl11}.

\subsubsection{Relationship to cognitive neuroscience}
SSMs were originally developed from a pure information processing
point of view and primarily used in psychology as a high-level
approximation of the decision process. More recent efforts in
cognitive neuroscience have simultaneously (i) validated core
assumptions of the model by showing that neurons indeed integrate
evidence probabilistically during decision making
\citep{SmithRatcliff04,GoldShadlen07} and (ii) applied this model to
understand and describe neural correlates of cognitive processes
\citep[e.g.][]{ForstmannAnwanderSchaferEtAl10,CavanaghWieckiCohenEtAl11}.\\

Multiple routes to decision threshold modulation have been identified. Decision threshold in the speed-accuracy trade-off is modulated by changes in the functional connectivity between pre-SMA and striatum \citep{ForstmannAnwanderSchaferEtAl10}. Neural network modeling \citep{Frank06,RatcliffFrank12} validated by studies of PD patients with a deep-brain-stimulator (DBS) in their subthalamic nucleus (STN) \citep{FrankSamantaMoustafaEtAl07} suggest that this node is implicated in raising the decision threshold when there is conflict between two options associated with similar rewards. This result was further corroborated by \citet{CavanaghWieckiCohenEtAl11} who found that frontal theta power (as measured by electroencelophagraphy and thought to correspond to conflict \citep{CavanaghZambrano-VazquezAllen12}) is correlated with decision threshold increase on a trial-by-trial basis. As predicted, this relationship was broken in PD patients with DBS turned on (but, critically, not when DBS was turned off thus showing the effect is not a result of the disease). In other words, by interfering with STN function through stimulation we were able to show that this brain area is causally involved in decision threshold modulation despite intact experience of conflict (as measured by theta power). Interestingly, these results provide a computational cognitive explanation for the clinical symptom of impulsivity observed in PD patients receiving DBS \citep{FrankSamantaMoustafaEtAl07}.

\subsubsection{Application to computational psychiatry}
Despite its long history, the DDM has only recently been applied to the study of psychopathology. For example, threat/no-threat categorization tasks (e.g. ``Is this word threatening or not?'') are used in anxiety research to explore biases to threat responses. Interestingly, participants with high anxiety are more likely to classify a word as threatening than low anxiety participants. One hypothesis assumes that this behavior results from an increased response bias towards threatening words in anxious people \citep{BeckerRinck04,Manguno-MireConstansGeer05,WindmannKruger98}. Using DDM analysis, \citet{White09} showed that instead of a response bias (or a shifted starting-point in DDM terminology), anxious people actually showed a perceptual bias towards classifying threatening words as indicated by an increased DDM drift-rate.\\

In a recent review article, \citet{WhiteRatcliffVaseyEtAl10} use this case-study to highlight the potential of the DDM to elucidate research into mental disease. Note that in this study the authors did not attempt to hypothesize about the underlying neural cause of this threat-bias. While there is some evidence that bias in decision making is correlated with activity in the parietial network \citep{ForstmannBrownDutilhEtAl10} this was not tested in respect to threatening words. Ultimately, I suggest that this research strategy should be applied to infer neural correlates of psychological DDM decision making parameters using functional methods like fMRI to the study neuropsychopathology (as outlined above).\\

The DDM has also been successfully used to show that ADHD subjects were less able to raise their decision threshold when accuracy demands were high \citep{MulderBosWeustenEtAl10}. Interestingly, the amount by which ADHD subjects failed to modulate their decision threshold correlated strongly with patients' impulsivity/hyperactivity rating. Moreover, this correlation was specific to impulsivity as no correlation was found between decision threshold modulation and inattentiveness.\\

In sum, SSMs show great promise as a tool for computational psychiatry. However, their applicability depend on the ability to accurately estimate them to construct individual CMDFs. In the following I will review different parameter estimation techniques. Special focus will be given to Bayesian methods. Finally, once SSMs can be fit accurately the question arises how to construct a map of mental illness based on these CMDF. Towards this goal I will review clustering methods that can be expressed in the Bayesian framework.

\subsection{Parameter estimation}
To identify computational biomarkers in a variable clinical population with the DDM it is critical to have robust and sensitive estimation methods. In the following I will describe traditional parameter estimation methods and their pitfalls. I will then argue how Bayesian estimation provides a complete framework that avoids these pitfalls.

\subsubsection{Maximum Likelihood and $\chi^2$}
Traditionally, fitting of computational models is treated as an optimization problem in which an objective function is minimized. Different objective functions for the DDM have been proposed. Most common is the quantile method which calculates the quantiles of the DDM likelihood and uses the $\chi^{2}$ statistic as the objective to compare it to the quantiles observed in the data \citep{RatcliffTuerlinckx02}. While this approach requires a fair amount of data to get stable quantile estimation it naturally deals with outlier RTs as the quantiles average extreme values out. In comparison, maximimum likelihood estimation (MLE) directly uses the DDM likelihood function derived from the DDM generative process \citep{NavarroFuss09}.\\

Psychological experiments often test multiple subjects on the same behavioral task. Models are then either fit to individual subjects or to the aggregated group data. Both approaches are not ideal. When models are fit to individual subjects we neglect any similarity the parameters are likely to have. While we do not necessarily have to make use of this property to make useful inference if we have lots of data, the ability to infer subject parameters based on the estimation of other subjects generally leads to more accurate parameter recovery \citep{SoferWieckiFrank} in cases where little data is available as is often the case in clinical and neurocoognitive experiments. One alternative is to aggregate all subject data into one meta-subject and estimate one set of parameters for the whole group. While useful in some settings, this approach is unsuited for the setting of computational psychiatry as individual differences play a huge role.

\subsubsection{Hierarchical Bayesian models}
Statistics and machine learning have developed efficient and versatile Bayesian methods to solve various inference problems \citep{Poirier06}. More recently, they have seen wider adoption in applied fields such as genetics \citep{StephensBalding09} and psychology \citep[e.g.][]{ClemensDeSelenEtAl11}. One reason for this Bayesian revolution is the ability to quantify the certainty one has in a particular estimation. Moreover, hierarchical Bayesian models provide an elegant solution to the problem of estimating parameters of individual subjects outlined above. Under the assumption that participants within each group are similar to each other, but not identical, a hierarchical model can be constructed where individual parameter estimates are constrained by group-level distributions \citep{NilssonRieskampWagenmakers11,ShiffrinLeeKim08}.

Bayesian methods require specification of a generative process in form of a likelihood function that produced the observed data $x$ given some parameters $\theta$. By specifying our prior belief we can use Bayes formula to invert the generative model and make inference on the probability of parameters $\theta$:

\begin{equation}\label{bayes}
P(\theta|x) = \frac{P(x|\theta) * P(\theta)}{P(x)}
\end{equation}

Where $P(x|\theta)$ is the likelihood and $P(\theta)$ is the prior probability. Computation of the marginal likelihood $P(x)$ requires integration (or summation in the discrete case) over the complete parameter space $\Theta$:

\begin{equation}
P(x) = \int_\Theta P(x|\theta) \, \mathrm{d}\theta
\end{equation}

Note that in most scenarios this integral is analytically intractable. Sampling methods like Markov-Chain Monte Carlo (MCMC) \citep{GamermanLopes06} circumvent this problem by providing a way to produce samples from the posterior distribution. These methods have been used with great success in many different scenarios \citep{GelmanCarlinSternEtAl03} and will be discussed in more detail below.

Another nice property of the Bayesian method is that it lends itself
naturally to a hierarchical design. In such a design, parameters for
one distribution can themselves come from a different distribution
which allows chaining together of distributions of arbitrary
complexity and map the structure of the data onto the model.

This hierarchical property has a particular benefit to cognitive modeling where data is often scarce. We can construct a hierarchical model to more adequately capture the likely similarity structure of our data. As above, observed data points of each subject $x_{i,j}$ (where $i = 1, \dots, S_j$ data points per subject and $j = 1, \dots, N$ for $N$ subjects) are distributed according to some likelihood function $f | \theta$.  We now assume that individual subject parameters $\theta_j$ are normal distributed around a group mean with a specific group variance ($\lambda = (\mu, \sigma)$ with hyperprior $G_0$) resulting in the following generative description:

\begin{align}\label{hierarchical}
  \mu, \sigma &\sim G_0() \\
  \theta_j &\sim \mathcal{N}(\mu, \sigma^2) \\
  x_{i, j} &\sim f(\theta_j)
\end{align}

\begin{figure}
\center
  \includegraphics{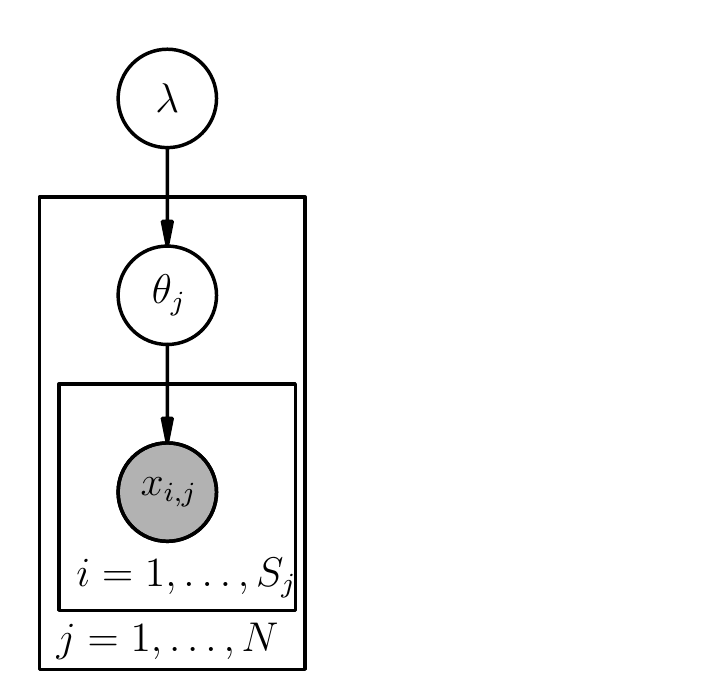}
\caption{Graphical notation of a hierarchical model. Circles represent continuous random variables. Arrows connecting circles specify conditional dependence between random variables. Shaded circles represent observed data. Finally, plates around graphical nodes mean that multiple identical, independent distributed random variables exist.}
\label{graphical_hierarchical}
  \end{figure}

See figure \ref{graphical_hierarchical} for the corresponding graphical model description.

Another way to look at this hierarchical model is to consider that our fixed prior on $\theta$ from formula (\ref{bayes}) is actually a random variable (in our case a normal distribution) parameterized by $\lambda$ which leads to the following posterior formulation:

\begin{equation}\label{hierarchical_posterior}
P(\theta, \lambda | x) = \frac{P(x|\theta) * P(\theta|\lambda) * P(\lambda)}{P(x)}
\end{equation}

Note that we can factorize $P(x|\theta)$ and $P(\theta|\lambda)$ due to their conditional independence. This formulation also makes apparent that the posterior contains estimation of the individual subject parameters $\theta_j$ and group parameters $\lambda$.

Several inference methods to estimate the posterior distribution have been developed. For details on commonly used sampling algorithms, see the appendix.\\

In sum, hierarchical Bayesian estimation leverages similarity between individual subject to share statistical power and increase sensitivity in our parameter estimation. However, note that in our computational psychiatry application the homogeneity assumption that all subjects come from the same normal distribution is almost certainly violated (see above). To deal with the heterogeneous data often encountered in psychiatry I will discuss mixture models further down below. Next, I will describe algorithms to estimate this posterior distribution.

\subsubsection{Likelihood-free methods}
Most models in cognitive neuroscience and mathematical psychology like the DDM are described by a latent generative process (see e.g. \eqref{wiener}). However, this generative description is usually ill suited for inference as it does not directly provide us with a closed-form likelihood function $p(x | \theta)$ of how observed data (e.g. the wiener first passage time; see above) arise from this generative process.\\

While the DDM is used partly because it has a tractable likelihood functions, many interesting considerations require models for which a generative process but no tractable likelihood can be specified. Recent examples of these efforts include changes of mind as new evidence is processed \citep{ResulajKianiWolpertEtAl09}, the influence of attention \citep{KrajbichRangel11}, and reward-based decision making given conflict in values of alternative actions \citep{CavanaghWieckiCohenEtAl11,RatcliffFrank12}. In these cases, a likelihood function must instead be approximated by simulation from the generative process. This type of inference is commonly called likelihood free.\\

It can be expected that the full spectrum of cognitive function will be relevant in computational psychiatry. Being able to formulate (and estimate) drift-diffusion models very flexibly potentially allows its application to new domains.

For a detailed explanation of these methods, see the appendix.

\subsection{Model selection}
Computational models often allow formulation of several plausible accounts of cognitive behavior. We anticipate that this problem will also occur in computational psychiatry where multiple theories of cognitive dysfunction must be tested. One way to differentiate between these various plausible hypothesis as expressed by alternative models is model comparison. In the following I will review various methods and metrics to compare hierarchical models. The most critical property for model comparison is that model complexity gets penalized because more complex models have greater explanatory power by design. Several model comparison measures have been devised. I refer the reader to the appendix section for mathematical details.

\subsection{Mixture Models}
In this section I will review different mixture models that allow estimation of clusters in data in a Bayesian framework. These are relevant to our objective as they (i) deal with the heterogeneity encountered in computational psychiatry and (ii) have the potential to bootstrap a new psychiatric classification system based on measurable, quantitative, computational endophenotypes. Because we are describing a toolbox using hierarchical Bayesian estimation techniques I will focus this section on mixture models as they are easily integrated into this framework. Where possible, I will highlight connections to more traditional clustering methods like k-means.

\subsubsection{Gaussian Mixture Models}
GMMs assume parameters to be distributed according to one of several Gaussian distributions (i.e. clusters). Specifically, given the number of clusters $k$, each cluster mean and variance gets estimated from the data.  This type of model is capable of solving our above identified problem of assuming heterogeneus subjects to be normally distributed: by allowing individual subject parameters to be assigned to different clusters we allow estimation of different sub-groups in our patient and healthy population. Note, however, that the number $k$ of how many clusters should be estimated must be specified a-priori in a GMM and remain fixed for the course of the estimation. This is problematic as we do not necessarily know how many sub-groups to expect in advance.  As we will see below, Bayesian non-parametrics solve this issue by inferring the number of clusters from data.

\subsubsection{Dirichlet Process Gaussian Mixture Models}
Dirichlet processes Gaussian mixture models (DPGMMs) belong to the class of Bayesian non-parametrics \citep{Antoniak74}. They can be viewed as a variant of GMMs with the critical difference that they assume an infinite number of potential mixture components (see \citet{GershmanBlei12} for a review).\\

Mixture models can infer sub-groups when the data is heterogeneous as is generally the case in patient populations. While the mindset describing these methods was their application towards the SSM their applicability is much more general than that. For example, the case-studies described above which used, among others, RL models to identify differences between HC and psychiatric patients could easily be embedded into this hierarchical Bayesian mixture model framework I outlined here. There are multiple benefits to such an approach. First, computational models fitted via hierarchical Bayesian estimation provide a tool to accurately describe the neurocognitive functional profile of individuals. Second, the mixture model approach is ideally suited to deal with the heterogeneity in patients but also healthy controls \citep{FairBathulaNikolasEtAl12}. Third, by testing psychiatric patients with a range of diagnoses (as opposed to most previous research studies that only compare patients with a single diagnosis, e.g. SZ, to controls) we might be able to identify shared pathogenic cascades as suggested by \citet{BuckholtzMeyer-Lindenberg12}.\\

\section{Conclusions}
As outlined above, computational psychiatry is an emerging field that shows great promise to understand aberrant biological processes in mental disease and address current challenges encountered in mental health research. By fitting computational models to behavioral data we can construct computational multi-dimensional features to replace symptom-based classification as implemented by the DSM. Decision making appears to provide a good framework for studying psychiatric disease as many disorders show abnormalities in core decision making processes. Sequential sampling models have a good track record in describing individual differences and can be linked to neuronal processes. Hierarchical Bayesian estimation provides a compelling toolbox to fit these models directly to data as it (i) provides an uncertainty measure; (ii) allows estimation of individual and group-level parameters simultaneously; (iii) allows for direct model comparison; (iv) can be used in scenarios where a likelihood can not be easily formulated; and (v) enable deconstruction of symptoms by identifying latent clusters. For example, impulsivity is a core symptom of impulse control disorders like ADHD, OCD, Tourette syndrome, substance abuse and eating disorder \citep{RobbinsGillanSmithEtAl12}. Computational cognitive models have already started to deconstruct this broadly defined behavioral symptom and identified separate pathways that can all lead to alterations in impulse control \citep{DalleyEverittRobbins11} including reduced motor inhibition \citep{ChamberlainFinebergBlackwellEtAl06,ChamberlainMenziesHampshireEtAl08} early temporal discounting of future rewards, insensitivity towards negative relative to positive outcomes \citep{FrankSantamariaOReillyEtAl07,CockburnHolroyd10}, or an inability to adjust the decision threshold appropriately \citep{MulderBosWeustenEtAl10,CavanaghWieckiCohenEtAl11,FrankSamantaMoustafaEtAl07}. Ultimately, the hope is to find a way to describe and diagnose psychiatric disease based on objective computational neurocognitive markers rather than the current subjective symptom-based approach. I believe that this combination of computational tools described here-within is powerful enough to lead the charge towards a new level of understanding of mental illness based on identifiable and reproducible neurocognitive computational multi-dimensional features.\\

\section{Appendix}
The following serves as a reference for the mathematical details of the methods motivated above.

\subsection{Drift-Diffusion Model}
\label{ddm_details}
Mathematically, the DDM is defined by a stochastic differential equation called the Wiener process with drift:
\begin{equation}\label{wiener}
  dW \sim \mathcal{N}(v, \sigma^2)
\end{equation}
where $v$ represents the drift-rate and $\sigma$ the variance. As we often only observe the response times of subjects we are interested in the wiener first passage time (wfpt) -- the time it takes $W$ to cross one of two boundaries. Assuming two absorbing boundaries of this process and through some fairly sophisticated math \citep[see e.g.][]{Smith00} it is possible to analytically derive the time this process will first pass one of the two boundaries (i.e. the wiener first passage time; wfpt). This probability distribution\footnote{the wfpt will not be a distribution rather than a single value because of the stochasticity of the wiener process} then serves as the likelihood function for the DDM. \\

\subsection{Bayesian Inference}
\label{inference_details}
\subsubsection{Empirical Bayesian Approximation}
Empirical Bayes can be regarded as an approximation of equation \eqref{hierarchical_posterior}. To derive this approximation consider $P(\theta | x)$ which we can calculate by integrating over $P(\lambda)$:

\begin{equation}
P(\theta | x) = \frac{P(x|\theta)}{P(x)} \int P(\theta|\lambda) P(\lambda) \,\mathrm{d}\lambda
\end{equation}

Now, if the true distribution $P(\theta|\lambda)$ is sharply peaked, the integral can be replaced with the point estimate of its peak $\lambda^\star$:

\begin{equation}
P(\theta | x) \simeq \frac{P(x|\theta) P(\theta|\lambda^\star)}{P(x|\lambda^\star)}
\end{equation}

Note, however, that $\lambda^\star$ depends itself on $P(\theta | x)$. One algorithm to solve this interdependence is Expectation Maximization (EM) \citep{DempsterLairdRubin77}. EM is an iterative algorithm that alternates between computing the expectation of $P(\theta | x)$ (this can be easily done by Laplace Approximation \citep{Azevedo-FilhoShachter94}) and then maximizing the prior point estimate $\lambda^\star$ based on the current values obtained by the expectation step. This updated point estimate is then used in turn to recompute the expectation. The algorithm is run until convergence or some other criterion in reached. This approach is used for example by \citet{HuysEshelONionsEtAl12} to fit their reinforcement learning models.

\subsubsection{Markov-Chain Monte-Carlo}
As mentioned above, the posterior is often intractable to compute analytically. While Empirical Bayes provides a useful approximation, an alternative approach is to estimate the full posterior by drawing samples from it. One way to achieve this is to construct a Markov-Chain that has the same equilibrium distribution as the posterior \citep{GamermanLopes06}. Algorithms of this class are called Markov-Chain Monte Carlo (MCMC) samplers.\\

One common and widely applicable algorithm is Metropolis-Hastings \citep{ChibGreenberg95,AndrieuFreitasDoucetEtAl03}. Assume we wanted to generate samples $\theta$ from the posterior $p(\theta | x)$. In general, we can not sample from $p(\theta | x)$ directly. Metropolis-Hastings instead generates samples $\theta^t$ from a proposal distribution $q(\theta^t|\theta^{t-1})$ where the next position $\theta^t$ only depends on the previous position at $\theta^{t-1}$ (i.e. the Markov-property). For simplicity we will assume that this proposal distribution is symmetrical; i.e. $q(\theta^t|\theta^{t-1}) = q(\theta^{t-1}|\theta^t)$. A common choice for the proposal distribution is the Normal distribution, formally:

\begin{equation}
\theta^t \sim \mathcal{N}(\theta^{t-1}, \sigma^2)
\end{equation}

The proposed jump to $\theta^t$ is then accepted with probability $\alpha$:

\begin{equation}\label{proposal}
  \alpha = \min (1, \frac{p(\theta^t | x)}{p(\theta^{t-1} | x)})
\end{equation}
In other words, the probability of accepting a jump depends on the probability ratio of the proposed jump position $\theta^t$ to the previous position $\theta^{t-1}$. Critically, in this probability ratio, the intractable integral in the denominator (i.e. $p(x) = \int p(x|\theta) \,\mathrm{d}\theta$) cancels out. This can be seen by applying Bayes formula \eqref{bayes}:
\begin{equation}
\frac{p(\theta^t | x)}{p(\theta^{t-1} | x)} =
\frac{
  \frac{p(x | \theta^t)p(\theta^t)}{p(x)}
}{
  \frac{p(x | \theta^{t-1})p(\theta^{t-1})}{p(x)}
}
=
\frac{p(x | \theta^t)p(\theta^t)}{p(x | \theta^{t-1})p(\theta^{t-1})}
\end{equation}

Thus, to calculate the probability of accepting a jump we only have to evaluate the likelihood and prior, {\it not} the intractable posterior.

Note that $\theta^0$ has to be initialized at some position and can not directly be sampled from the posterior. From this initial position, the Markov chain will explore other parts of the parameter space and only gradually approach the posterior region. The first samples generated are thus not from the true posterior and are often discarded as ``burn-in''. Note moreover that once the algorithm reaches a region of high probability it will continue to explore lower probability regions in the posterior, albeit with lower frequency. This random-walk behavior is due to the probability ratio $\alpha$ which allows Metropolis-Hastings to also sometimes accept jumps from a high probability position to a low probability position.\\

Another common algorithm is Gibbs sampling that iteratively updates each individual random variable conditional on the other random variables set to their last sampled value \citep[e.g][]{FreyJojic05}. Starting at some configuration $\theta^0$, the algorithm makes $T$ iterations over each random variable $\theta_i$. At each iteration $t$ each random variable is sampled conditional on the current ($t-1$) value of all other random variables that it depends on:

\begin{equation}
 \theta_i^{t} \sim p(\theta_i^{(t)}|\theta_{i\neq j}^{(t-1)})
\end{equation}

Critically, $\theta_{i\neq j}^{(t-1)}$ are treated as constant. The sampled value of $\theta_i^{(t)}$ will then be treated as fixed while sampling the other random variables.

Note that while Gibbs sampling never rejects a sample (which often leads to faster convergence and better mixing), in contrast to Metropolis-Hastings, it does require sampling from the conditional distribution which is not always tractable.

\subsection{Likelihood free methods}
Several likelihood-free methods have emerged in the past (for a review, see \citet{TurnerVan12}). Instead of an analytical solution of the likelihood function, these methods require a sampling process that can simulate a set of data points from a generative model for each $\theta$. We will call the simulated data $y$ and the observed data $x$. Approximate Bayesian Computation (ABC) relies on a distance measure $\rho(x, y)$ that compares how similar the simulated data $y$ is to the observed data $x$ (commonly, this distance measure relies on summary statistics). We can then use the Metropolis-Hastings algorithm introduced before and change the acceptance ration $\alpha$ \eqref{proposal} to use $\rho(x, y)$ instead of a likelihood function.
\begin{equation}
  \alpha = \left\{
    \begin{array}{l l}
    \min (1,\frac{p(\theta^t)}{p(\theta^{t-1})}) & \quad \text{if $\rho(x, y) \leq \epsilon_0$} \\
    0 & \quad \text{if $\rho(x, y) \geq \epsilon_0$}
    \end{array}
    \right.
\end{equation}

where $\epsilon_0$ is an acceptance threshold. Large $\epsilon_0$ will result in higher proposal acceptance probability but a worse estimation of the posterior while small $\epsilon_0$ will lead to better posterior estimation but slower convergence.\\

\begin{figure}
  \includegraphics[width=\columnwidth]{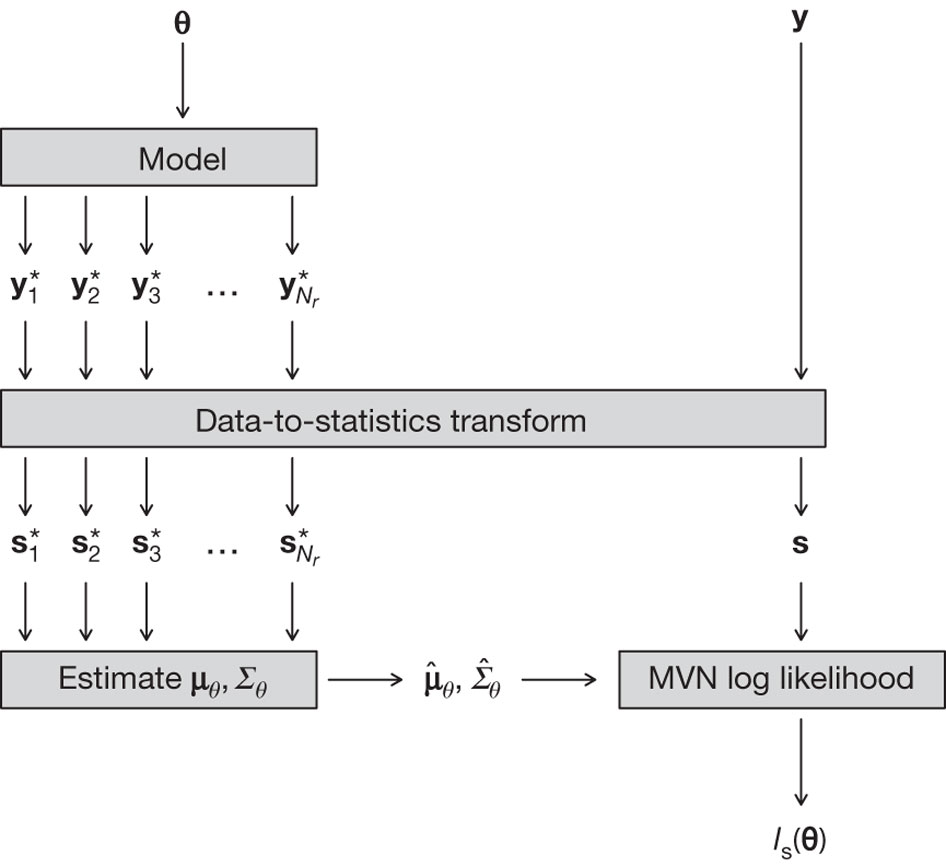}
  \caption{Construction of a synthetic likelihood. To evaluate parameter vector $\theta$, $N_r$ data sets $y_{1,\dots,N_r}$ are sampled from the generative model. On each sampled data set summary statistics $s_{1,\dots,N_r}$ are computed. Based on these summary statistics a multivariate normal distribution is approximated with mean $\mu_\theta$ and covariance matrix $\Sigma_\theta$. The likelihood is approximated by evaluating summary statistics of the actual data on the log normal distribution with the estimated $\mu_\theta$ and     $\Sigma_\theta$. Reproduced from \citep{Wood10}.}
  \label{synth_like}
\end{figure}

An alternative approach to ABC is to construct a synthetic likelihood function based on summary statistics \citep{Wood10}. Specifically, we sample $N_r$ multiple data sets $y_{1,\dots,N_r}$ from the generative process. We then compute summary statistics $s_{1,\dots,N_r}$ for each simulated data set\footnote{The summary statistics must (i) be sufficient and (ii) normally distributed}. Based on these summary statistics we then construct the synthetic likelihood function to evaluate $\theta$ (see figure \ref{synth_like} for an illustration):

\begin{equation}
  p(x | \theta) \simeq \mathcal{N}(S(x); \mu_{\theta}, \Sigma_{\theta})
\end{equation}

This synthetic likelihood function based on summary statistics can then be used as a drop-in replacement for e.g. the Metropolis-Hastings algorithm outlined above.

\subsection{Model Comparison}
\label{comparison_details}
\subsubsection{Deviance Information Criterion}
The Deviance Information Criterion (DIC) is a measure which trades off model complexity and model fit \citep{SpiegelhalterBestCarlin02}. Several similar measures exist such as Akaike Information Criterion (AIC) and the Bayesian Information Criterion (BIC). However, both these measures use the number of parameters as a proxy for model complexity. While a reasonable approximation to the complexity of non-hierarchical models, the relationship between model parameters (some of which are latent) and complexity in hierarchical models is more intricate. The DIC measure instead infers the number of parameters from the posterior. The DIC is computed as follows:

\begin{equation}
\text{DIC} = \bar{D} + pD
\end{equation}

where

\begin{equation}
pD = \bar{D} - \hat{D}
\end{equation}

$\bar{D}$ is the posterior mean of the deviance (i.e. $-2*\text{log}(\text{likelihood})$) and $\hat{D}$ is a point estimate of the deviance obtained by substituting in the posterior means. Loosely, $\bar{D}$ represents how well the model fits the data on average while $\hat{D}$ captures the deviance at the best fitting parameter combination. $pD$ then acts as a measure related to the posterior variability and used as a proxy for the effective number of parameters. Complex models with many parameters will tend to have higher posterior variability and thus result in increased $pD$ penalization.\\

Note that the only parameters that affect $\hat{D}$ directly in our hierarchical model (equation \ref{hierarchical_posterior}) are the subject parameters $\theta_i$. Thus, DIC estimates model fit based on how well individual subjects explain the observed data.

\subsubsection{BIC}
The Bayesian Information Criterion (BIC) is defined as follows:

\begin{equation}
\text{BIC} = -2 * \text{log}p(x|\hat{\theta}^{ML}) + k * \text{log}(n)
\end{equation}

where $k$ is the number of free parameters, $n$ is the number of data points, $x$ is the observed data and $\text{log} p(x|k)$ is the likelihood of the parameters given the data \citep{Schwarz78}.

While BIC can not directly be applied to hierarchical models (as outlined above), it is possible to integrate out individual subject parameters \citep[e.g.][]{HuysEshelONionsEtAl12}:

\begin{equation}
\text{log}p(x|\hat{\theta}^{ML}) = \sum_{i} \text{log} \int p(x_{i}|h) p(h|\hat{\theta}^{ML}) \, \mathrm{d}h
\end{equation}

where $x_{i}$ is the data belonging to the $i$th subject. The resulting score is called integrated BIC.\\

Since the subject parameters are integrated out, integrated BIC estimates how well the group parameters are able to explain the observed data.

\subsubsection{Bayes Factor}
Another measure to compare two models is the Bayes Factor (BF) \citep{KassRaftery95}. It is defined as the ratio between the marginal model probabilities of the two models:

\begin{equation}
BF = \frac{p(x|M_1)}{p(x|M_2)}
= \frac{\int p(\theta_1|M_1) p(x|\theta_1,M_1)\,\mathrm{d}\theta_1}
{\int p(\theta_2|M_2) p(x|\theta_2,M_2)\,\mathrm{d}\theta_2}
\end{equation}

The magnitude of this ratio informs the degree one should belief in one model compared to the other.\\

As BF integrates out subject {\it and} group parameters this model comparison measure should be used when different classes of models are to be compared in their capacity to explain observed data.

\subsection{Mixture Models}
\subsubsection{Gaussian Mixture Models}
\label{gmm_details}
Mixture models infer $k$ number of clusters in a data set. The assumption of normal distributed clusters leads to a Gaussian Mixture Model (GMM) with a probability density function as follows:
\begin{equation}
p(x | \pi, \mu_{1,\dots,K}, \sigma_{1,\dots,K}) = \sum_{k=1}^{K}\pi_k \mathcal{N}(x_i|\mu_k, \sigma^2_k)
\end{equation}

Each observed data point $x_i$ can be created by drawing a sample from the normal distribution selected by the unobserved indicator variable $z_i$ which itself is distributed according to a multinomial distribution $\pi$:
\begin{align}
  \mu_k, \sigma_k &\sim \operatorname{G_0}()\\
  z_i &\sim \pi \\
  x_i &\sim \mathcal{N}(\mu_{z_i}, \sigma^2_{z_i})
\end{align}

where the base measure $\operatorname{G_0}$ defines the prior for $\mu_k$ and $\sigma_k$. To simplify the inference it is often advisable to use a conjugate prior for these paramters. For example, the normal distribution is the conjugate prior for a normal distribution with known variance:

\begin{align}
  \mu_k &\sim \mathcal{N}(\mu_0, \sigma_0)
\end{align}

In a similar fashion, we can assign the mixture weights a symmetric Dirichlet prior:

\begin{align}
  \pi &\sim \operatorname{Dir}(\frac{\alpha}{K}, \dots, \frac{\alpha}{K})
\end{align}

Note that the GMM assumes a mixture distribution on the level of the observed data $x_i$. However, in our relevant case of a multi-level hierarchical model we need to place the mixture at the level of the latent subject parameters instead of the observed data. As before, we use the subject index $j = {1,\dots,N}$.

\begin{align}
  \mu_k, \sigma_k &\sim \operatorname{G_0}() \\
  \pi &\sim \operatorname{Dir}(\alpha) \\
  z_j &\sim \operatorname{Categorical}(\pi) \\
  \theta_{j} &\sim \mathcal{N}(\mu_{z_j}, \sigma^2_{z_j}) \\
   x_{i,j} &\sim \operatorname{f}(\theta_j)
\end{align}

Where $\operatorname{f}$ denotes the likelihood function.\\

Interestingly, the famous K-Means clustering algorithm is identical to a Gaussian Mixture Model (GMM) in the limit $\sigma^2 \rightarrow 0$ \citep{KulisJordanEecsEtAl12}. K-Means is an expectation maximization (EM) algorithm that alternates between an expectation step during which data points are assigned to their nearest cluster centroids and a maximization step during which new cluster centroids are estimated. This algorithm is repeated until convergence is reached (i.e. no points are reassigned to new clusters).\\

\subsubsection{Dirichlet Process Gaussian Mixture Models}
\label{dirichlet_details}

\begin{equation}
p(x | \pi, \mu_{1,\dots,\infty}, \sigma_{1,\dots,\infty}) = \sum_{k=1}^{\infty}\pi_k \mathcal{N}(x_i|\mu_k, \sigma^2_k)
\end{equation}

As above, we specify our generative mixture model:

\begin{align}
  \mu_k, \sigma_k &\sim \operatorname{G_0}() \\
  z_i &\sim \operatorname{Categorical}(\pi) \\
  x_i &\sim \mathcal{N}(\mu_{z_i}, \sigma^2_{z_i})
\end{align}

with the critical difference of replacing the hyperprior $\pi$ with the {\it stick breaking process} \citep{Sethuraman91}:
\begin{align}
  \pi &\sim \operatorname{StickBreaking}(\alpha)
\end{align}

\begin{figure}
\includegraphics{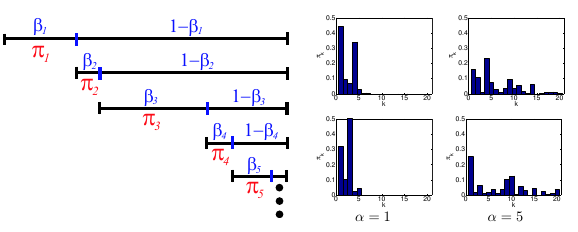}
\caption{Left: Stick-breaking process. At each iteration (starting from the top) a $\pi$ is broken off with relative length $\sim \operatorname{Beta}(1, \alpha)$. Right: Histogram over different realizations of the stick-breaking process. As can be seen, higher values of hyperprior $\alpha$ lead to a more spread out distribution. Taken from Eric Sudderth's PhD thesis.}
\label{stick_breaking}
\end{figure}

The stick-breaking process is a realization of a Dirichlet process (DP). Specifically, $\pi = \left\{\pi_k\right\}_{k=1}^{\infty}$ is an infinite sequence of mixture weights derived from the following process:

\begin{align}
 \beta_k &\sim \operatorname{Beta}(1, \alpha) \\
 \pi_k &\sim \beta_k * \prod_{l=1}^{k-1} (1-\beta_l)
\end{align}

with $\alpha > 0$. See figure \ref{stick_breaking} for a visual explanation.\\

\begin{figure}
\includegraphics[width=\columnwidth]{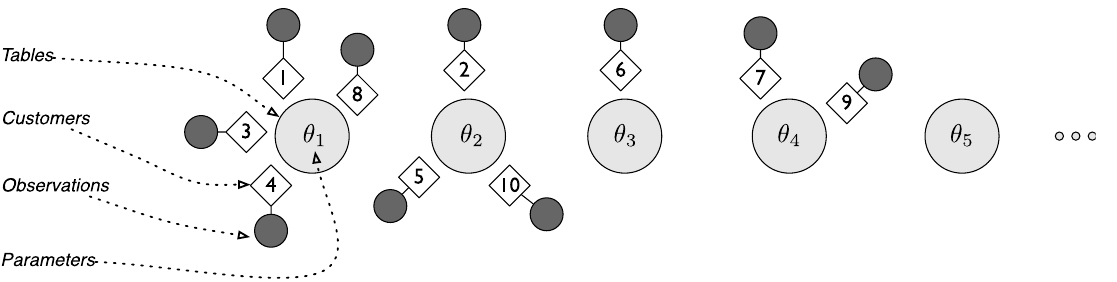}
\caption{Illustration of the Chinese Restaurant Process. Customers are seated at tables with parameters $\theta$. The more customers are already seated at a table, the higher the probability that future customers are seated at the same table (i.e. clustering property). Taken from \citet{GershmanBlei12}.}
\label{crp}
\end{figure}

The Chinese Restaurant Process (CRP) -- named after the apparent infinite seating capacity in Chinese restaurants -- allows for a more succinct model formulation. Consider that customers $z_i$ are coming into the restaurant and are seated at table $k$ with probability:

\[ p(z_i = k|z_{1,\dots,n-1},\alpha,K) = \frac{n_k + \alpha/K}{n- 1 + \alpha} \]

where $k = 1 \dots K$ is the table and $n_k$ is the number of customers already sitting at table $k$ (see figure \ref{crp} for an illustration). It can be seen that in the limit as $K\to\infty$ this expression becomes:

\[ p(z_i = k|z_{1,\dots,n-1},\alpha) = \frac{n_k}{n - 1 + \alpha} \]

Thus, as customers are social, the probability of seating customer $z_i$ to table $k$ is proportional the number of customers already sitting at that table. This desirable clustering property is also known as the ``rich get richer''.\\

Note that for an individual empty table $k$ at which no customer has been seated (i.e. $n_k = 0$) the probability of seating a new customer to that table goes to 0 in the limit as $K\to\infty$. However, at the same time the number of empty tables approaches infinity. Consider that we have so far seated $L$ customers to tables and the set $\mathbf{Q}$ contains all empty tables such that there are $|\mathbf{Q}| = K - L$ empty tables in the restaurant. The probability of seating a customer $z_i$ at an empty table becomes:

\[ p(z_i \in \mathbf{Q}|\mathbf{z}_{1,\dots,n-1},\alpha)
 =  \frac{\alpha}{n-1+\alpha} \]

As can be seen, the probability of starting a new table is proportional to the concentration parameter $\alpha$. Intuitively, large values of the dispersion parameter $\alpha$ lead to more clusters being used.\\

Thus, while the Stick-Breaking process sampled mixture weights from which we had to infer cluster assignments, the CRP allows for direct sampling of cluster assignments. The resulting model can then be written as:

\begin{align}
  \mu_k, \sigma_k &\sim \operatorname{G_0}() \\
z_{1,\dots,N} &\sim \operatorname{CRP}(\alpha) \\
x_i &\sim \mathcal{N}(\mu_{z_i}, \sigma^2_{z_i})
\end{align}

Finally, in a hierarchical group model we would need to place the infinite mixture on the subject level rather than the observed data level:
\begin{align}
  \mu_k, \sigma_k &\sim \operatorname{G_0}() \\
  z_j &\sim \operatorname{CRP}(\alpha) \\
  \theta_{j} &\sim \mathcal{N}(\mu_{z_j}, \sigma^2_{z_j}) \\
  x_{i,j} &\sim \operatorname{F}(\theta_j)
\end{align}

\begin{figure}
  \center
\includegraphics{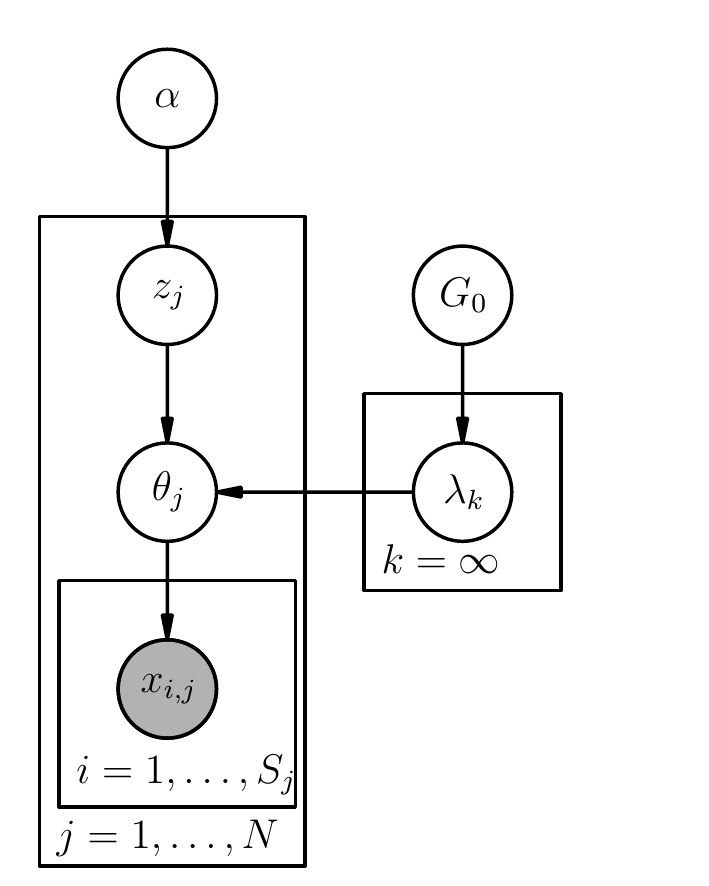}
\caption{Graphical model representation of the hierarchical Dirichlet process mixture model. Group parameters $\lambda_k = (\mu_k, \sigma_k)$. See text for details.}
\label{graphical_model_hierarchical_dpgmm}
\end{figure}

See figure \ref{graphical_model_hierarchical_dpgmm} for a graphical model description.

Note that while the potential number of clusters is infinite, any realization of this process will always lead to a finite number of clusters as we always have finite amounts of data. However, this method allows the addition (or subtraction) of new clusters as new data becomes available.

\bibliographystyle{plainnat}
\bibliography{ccnlab,ccnlab.ip,mendeley,Whyking,Whyking_formatted}
\end{document}